\documentclass[conference]{IEEEtran}
\IEEEoverridecommandlockouts
\usepackage{cite}
\usepackage{amsmath,amssymb,amsfonts}
\usepackage[linesnumbered,ruled,vlined,commentsnumbered]{algorithm2e}
\usepackage{graphicx}
\usepackage{textcomp}
\usepackage{xcolor,bm,subfigure,multirow}
\usepackage{graphicx,subfigure}
\usepackage{amsmath}
\usepackage{array}
\usepackage{multirow}
\usepackage{amssymb}
\usepackage{dblfloatfix}
\usepackage{makecell}
\usepackage{bm}
\usepackage{color}
\usepackage{float}
\usepackage{multirow}
\usepackage[margin=0.7in]{geometry}

\def\BibTeX{{\rm B\kern-.05em{\sc i\kern-.025em b}\kern-.08em
T\kern-.1667em\lower.7ex\hbox{E}\kern-.125emX}}

\begin{document}

\title{Multi-Task Offloading over Vehicular Clouds under Graph-based Representation}

\author{\IEEEauthorblockN{Minghui LiWang\IEEEauthorrefmark{1}\IEEEauthorrefmark{2}, Zhibin Gao\IEEEauthorrefmark{1}, Seyyedali Hosseinalipour\IEEEauthorrefmark{3}, and Huaiyu Dai\IEEEauthorrefmark{3}}
\IEEEauthorblockA{\IEEEauthorrefmark{1}\textit{Dept. of Information and Communication Engineering,}
\textit{Xiamen University,} Xiamen, China \\}
\IEEEauthorblockA{\IEEEauthorrefmark{2}\textit{Dept. of Electrical and Computer Engineering,}
\textit{University of Western Ontario,} London, Canada\\}
\IEEEauthorblockA{\IEEEauthorrefmark{3}\textit{Dept. of Electrical and Computer Engineering,}
\textit{North Carolina State University,} Raleigh, USA \\
Email: \{minghuilw@stu.xmu.edu.cn, gaozhibin@xmu.edu.cn, shossei3@ncsu.edu, hdai@ncsu.edu\}}

%\thanks{This work is supported in part by the National Natural Science Foundation
%of China (grant nos. 61971365, 61871339), Digital Fujian Province Key Laboratory
%of IoT Communication, Architecture and Safety Technology (grant no. 2010499), the
%US National Science Foundation (grant nos. ECCS-1444009, CNS-1824518), the Major
%Research Plan of the National Natural Science Foundation of China (grant no. 91638204),
%and the State Key Program of the National Natural Science Foundation of China (grant
%no. 61731012).
%}
}

\maketitle

\begin{abstract}
Vehicular cloud computing has emerged as a promising paradigm for realizing
user requirements in computation-intensive tasks in modern driving environments.
In this paper, a novel framework of multi-task offloading over vehicular clouds (VCs)
is introduced where tasks and VCs are modeled as undirected weighted graphs. Aiming
to achieve a trade-off between minimizing task completion time and data exchange
costs, task components are efficiently mapped to available virtual machines in the
related VCs. The problem is formulated as a non-linear integer programming problem,
mainly under constraints of limited contact between vehicles as well as available resources,
and addressed in low-traffic and rush-hour scenarios. In low-traffic cases, we determine
optimal solutions; in rush-hour cases, a connection-restricted random-matching-based
subgraph isomorphism algorithm is proposed that presents low computational complexity.
Evaluations of the proposed algorithms against greedy-based baseline methods are
conducted via extensive simulations.
\end{abstract}


\section{Introduction}

The Internet of vehicles (IoV) is an emerging paradigm that
enables innovations related to immersive vehicular applications such as autonomous driving and advanced driver assistants, offering safety, 
convenience and entertainment experiences for drivers and passengers~\cite{1}. Moreover, technological advances in computing processors and sensing devices facilitate tasks with innovative and computation-intensive features (e.g., self-driving and simultaneous localization and mapping), which require
massive computational resources. Graph-based representation can be used to characterize
some of these tasks: each task is
modeled as a graph, where the vertices (components)\footnote{``Component''
is used instead of ``vertex'' throughout this paper to
convey the physical significance.} represent either data sources or
data processing units while edges describe the dependency (data flows) between the vertices~\cite{2,3,4}.
However, the computational resources and capability limitations of on-board equipments~\cite{4,5} pose
major challenges to IoV, such that the inherent limitation of a single smart vehicle
may hinder fulfillment of task execution requirements. Vehicular cloud computing
(VCC) has thus been proposed, wherein vehicles act as computing servers to form vehicular
clouds (VCs) by sharing surplus resources with users facing heavy on-board workloads
via opportunistic vehicle-to-vehicle (V2V) communications~\cite{5}.

The literature on computation-intensive task offloading can be roughly
divided into two categories: 1) tasks that are directly mapped as bit streams and
considered a collection of sub-tasks without considering inherent dependencies~\cite{6,7,8,9};
and 2) tasks that are modeled as directed/undirected graphs by considering the inner
dependencies of sub-tasks, such as~\cite{2,3,4,10,11,12,13}. Furthermore, graph-based task offloading can be classified into three
types: static, semi-static, and dynamic, according to the features of network
topology. Assuming static topologies of servers and users in the cloud computing context,
a randomized scheduling algorithm is proposed in~\cite{4}, which stabilizes a system with graph job arrivals/departures and facilitates a smooth trade-off between minimizing the average execution
cost and queue length.

For the semi-static environment where the
locations of either servers or users are fixed, in~\cite{2}, applications were
modeled as directed tasks, and sequential and concurrent task offloading mechanisms
were presented to minimize application completion time. The authors
in~\cite{3} put forth a novel framework for energy-efficient graph job allocation
in geo-distributed cloud networks, where solutions were provided for data center
networks of varying scales. A Lyapunov-optimization-based dynamic offloading approach
for directed-graph-based jobs was described in~\cite{10}. The proposed approach
satisfies the constraints on both energy conservation and application execution time.
In~\cite{11}, scheduling of parallel jobs composed of a set of independent tasks
was studied by considering energy consumption and job completion time. A fast hybrid
multi-site computation offloading mechanism was proposed in~\cite{12}, where
offloading solutions considering application sizes were obtained in a timely manner.
Different from the abovementioned scenarios, a VC-based computation offloading mechanism
was studied in~\cite{13} where computing missions were modeled as tasks with
interdependency and executed in different vehicles to minimize overall response time
while enhancing the capacity of users (edge clouds).

Graph-based task offloading in dynamic network environments have rarely been investigated up to now, where the
mobility of servers and users as well as the interdependency of components pose challenges
to the design of offloading mechanisms. Consequently, approaches in static network
systems are difficult to implement in dynamic environments. In our previous study~\cite{19},
a randomized graph job allocation mechanism over VCs based on hierarchical tree
decomposition was proposed. This paper is regarded as follow-up work of~\cite{19}, in
which multi-task offloading and various attributes of components and service providers
(SPs) are considered; such factors present additional challenges related to the problem
size and algorithm efficiency. Moreover, potential competition for resources between
components must be considered due to task concurrency. To the best of our knowledge,
we are among the first to examine the multi-task offloading problem over
VCs while considering their inherent characteristics.

In this paper, we introduce a novel VC-assisted computation offloading
framework that captures a multi-server and multi-user environment, where each user
(task owner (TO)) has a task modeled as an undirected weighted graph with specific
requirements for execution time. Virtual machine (VM)-based~\cite{14} representation is utilized 
to quantify available resources of vehicles (SPs) in the VC. The VC is abstracted
as an undirected weighted graph of SPs with VMs that can provide heterogeneous computational
capabilities. Through the proposed mechanism,
each component is efficiently mapped (offloaded) to an appropriate VM under constraints
of opportunistic communications and available resources. The main contributions of
this paper are as follows:

\begin{itemize}
\item A novel VC-assisted computation offloading framework is presented that accounts
for concurrent multi-tasks under graph-based representation. Our framework alleviates heavy TO workloads caused by limitations in on-board resources and computational capabilities.

\item To
achieve the trade-off between minimizing the task completion time and data exchange costs, the graph-based concurrent multi-task offloading problem is formulated as a nonlinear integer
programming (NIP) problem under limited opportunistic contact between SPs and
available VMs.

\item To tackle the aforementioned NIP problem, an optimal algorithm is first
developed to find the best offloading solution in low-traffic scenarios. This approach
relies on addressing the subgraph isomorphism problem\footnote{The subgraph isomorphism
problem is a task where two graphs, $H_1$ and $H_2$, are given as the input,
and one must determine whether $H_1$ contains a subgraph that is isomorphic
to $H_2$.}, which is known to be NP-complete~\cite{3,18}. The optimal algorithm is thus ineffective when encountering a large number of tasks
and SPs or when facing complicated task and network topologies. To address this issue,
a connection-restricted random matching algorithm is proposed, which exhibits low computational complexity and works well for larger
and fast-changing IoV networks.

\item Two greedy-based methods are proposed to serve as baselines for performance evaluation. We demonstrate that the proposed connection-restricted random matching algorithm achieves a similar performance to the optimal algorithm with a significantly lower running time while outperforming
baseline methods under various scenarios.
\end{itemize}
The rest of this paper is organized as follows. The system models and problem formulation are presented in Section II. The optimal and the connection-restricted random matching algorithms are presented in Section III and IV, respectively. In Section V, the performance evaluation through comprehensive simulations is introduced before conclusion in Section VI.

\section{System model and problem formulation}

In this paper, a novel computation offloading framework is proposed where
multi-tasks can be mapped to SPs in the relevant VC via one-hop V2V interactions.
SPs own different quantities of available VMs, which can process task components
by providing various computational capabilities measured by execution time. Note
that structural characteristics exist in computation-intensive tasks and IoV topology,
due to which the tasks of TOs and the VC network are both modeled as weighted undirected
graphs. Under constraints of limited opportunistic vehicular contact duration and
available resources, each VC aims to effectively allocate all of the task components to
SPs while achieving a trade-off between task completion time and data exchange
costs. Assuming that a VC contains SP set $\bm S$ and TO set $\bm O$, the related models are introduced below.

\subsection{Vehicular contact model}

The interaction between vehicles $x$ and $y$ during $t\in ({\tau}_{1}, {\tau}_{2})$ occurs when the
following conditions are satisfied: $\left\| L_x\left({\tau}_{1}\right)-L_y
({\tau}_{1})\right\|>R, \left\| L_x
(t)-L_y (t) \right\|\le R, $ and $\left\|L_x
\left({\tau}_{2}\right)-L_y ({\tau}_{2})
\right\|>R$, where $L_x (\tau) $ and $L_y (\tau) $ denote vehicles' locations at time $\tau$, and $\|\cdot \| $ and $R$ represent
the Euclidean distance and the vehicular communication radius, respectively.
Generally,
the contact duration between vehicle $x$ and $y$ obeys an exponential distribution~\cite{15,16} with parameter $\lambda_{xy}$; therefore, the probability that the contact duration $\Delta
\tau_{xy}$ between vehicles $x$ and $y$ is larger than $T$ is given
by $P\left(\Delta{\tau}_{xy}>T, \lambda_{xy}\right)=e^{-T\lambda_{xy}}$.
The larger the value of $e^{-T\lambda_{xy}}$, the more assurance can be achieved to protect the required
data interaction between moving vehicles.

%f1
\begin{figure*}[h!t]
\centering
\subfigure[]{\includegraphics[width=3.35in,height=5.55cm]{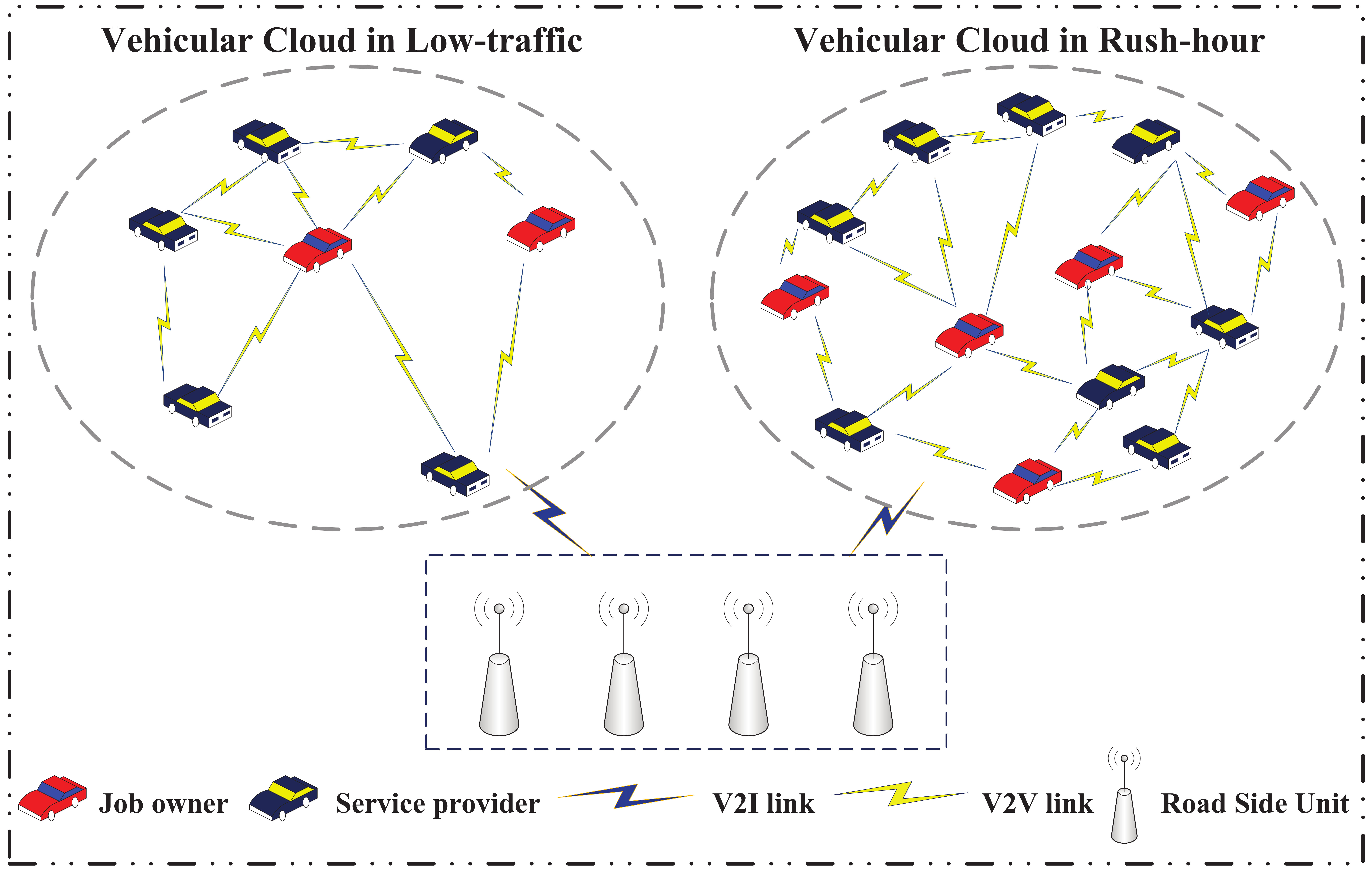}}\quad
\subfigure[]{\includegraphics[width=3.35in,height=5.55cm]{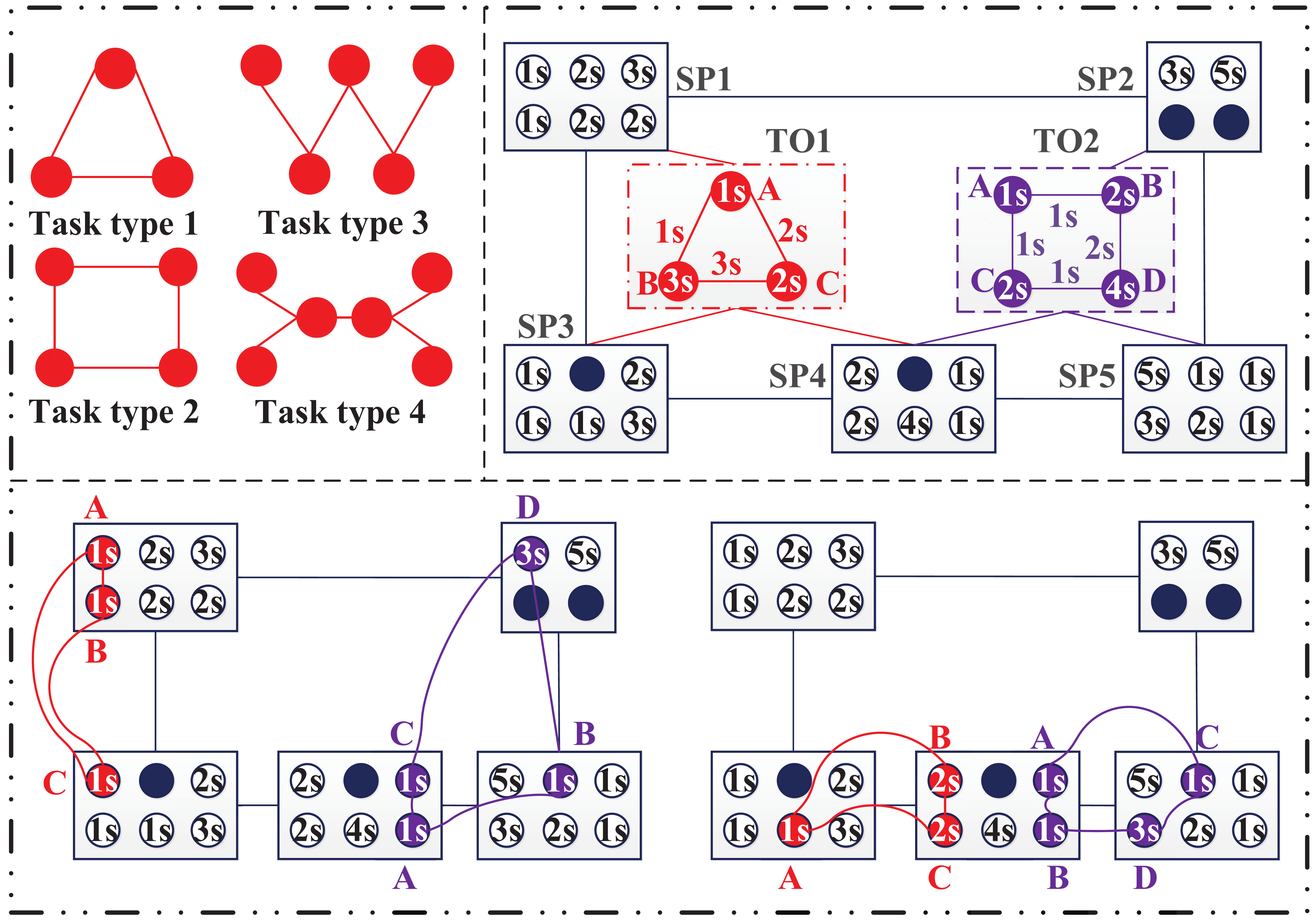}}
\caption{(a) System model of a vehicular cloud network; (b) Several task types and examples of multi-task offloading solution under graph-based representation.}
\label{fig1}
\end{figure*}

\subsection{Vehicular cloud modeling via graphs}

Each SP $s_y\in \bm{S}$ has a collection of idle VMs ${\bm{m}}_{
\bm{y}}$ that are fully connected, with various computational
capabilities related to the execution time for processing one component. Moreover,
each available VM can run only one component at a time. A VC covers a region containing
several TOs and SPs, where at least one SP exists in each TO's communication range.
It is assumed that TOs are under consistent pressure due to insufficient local resources,
such that TOs prefer to integrate resources from SPs in the related VC; thus, we
do not consider VMs on TOs as available resources. Consequently, a VC is represented as a graph ${\bm{G}}^{\bm{s}}=({
\bm{V}}^{\bm{s}}, {\bm{E}}^{\bm{s}}, {\bm{W}}^{
\bm{s}})$ containing a set of SPs ${\bm{V}}^{\bm{s}}
=\{s_y|s_y\in \bm{S}\}$, in which each SP ${s}_y$ has a collection of available VMs ${\bm{m}}_{\bm{y}}=
\{s^{y}_{j}|j\in \{1,2,\dots, \left|{\bm{m}}_{\bm{y}}\right|\}\}$, where $s^{y}_{j}$ denotes
the ${}j^{\rm th}$ VM of $s_y$; and the related computational capability set $\{t^{s_y}_j|j
\in \{1,2,\dots, \left|{\bm{m}}_{
\bm{y}}\right|\}\}$, where $t^{s_y}_j$ represents the execution time that VM $s^{y}_{j}$ can provide to process one component in a graph job. The edge set ${\bm{E}}^{
\bm{s}}=\{e^s_{yy'}|s_y\in
\bm{S}, s_{y'}\in \bm{S}, y\ne y'\}$, where $e^s_{yy'}$ indicates that $s_y$ can
communicate with $s_{y'}$ via a one-hop V2V channel; and the weight of
the edge set ${\bm{W}}^{\bm{s}}=\{{\lambda
}_{yy'}|s_y\in \bm{S}, s_{y'}
\in \bm{S}, y\ne y'\}$ describes the mean of the
corresponding parameters of the exponential distribution of contact duration
between vehicles.

\subsection{Task modeling via graphs}

The task of a TO ${o}_x\in \bm O$ is depicted as a graph ${\bm{G}}^{
\bm{o_x}}=({\bm{V}}^{\bm{o_x}}, $ ${
\bm{E}}^{\bm{o_x}}, {\bm{W}}^{\bm{o_x}})$, which contains a set of components ${\bm{V}}^{\bm{o_x}} =\{v^{o_x}_{i}|i\in {\bm{n}}_{\bm{x}}\}$ where ${\bm{n}}_{\bm{x}}=\{{1,...,|{\bm{V}}^{\bm{o_x}}|}\}$; here,
each component $v^{o_x}_{i}$ has the maximum tolerant execution time $t^{o_x}_i $ (seconds)\footnote{It is assumed that task components are of the same data size, but each
component requires different computational capabilities, which is described by the maximum tolerant execution time (e.g., component $v^{o_x}_i$ should
be completed within $t^{o_x}_i$) seconds.} and a set of edges ${\bm{E}}^{
\bm{o_x}}=\{e^{o_x}_{ii'}|i\in {\bm{n}}_{\bm{x}},
i'\in {\bm{n}}_{\bm{x}}, i\ne i'\}$ with associated
weights ${\bm{W}}^{\bm{o_x}}$ $=$ $\{{\omega
} ^{o_x}_{ii'}|i\in {\bm{n}}_{\bm{x}}, i'\in {\bm{n}}_{
\bm{x}}, i\ne i'\}$. Graph ${\bm{G}}^{\bm{o_x}}$ represents how the computation should be split among components
in ${\bm{V}}^{\bm{o_x}}$. Specifically, edges represent required data flows
between components, and the weight ${\omega}^{o_x}_{ii'}$ of edge $e^{o_x}_{ii'}$ indicates
the requested connecting duration
between components $v^{o_x}_{i}$ and $v^{o_x}_{i'}$, which is considered to be the lowest execution time between these components. Also, the contact duration of SPs that
handle these components should be equal to or greater
than ${\omega}^{o_x}_{ii'}.$ For example, if two connected components $v^{o_1}_1$ and $v^{o_1}_2$ are offloaded to different VMs $s^1_1$ and $s^3_2$ that can provide execution time 1.5 and 3 seconds, respectively. Then, 
${\omega}^{o_1}_{12}=min(1.5, 3)=1.5$ seconds are required for the intermediate data collection and transmission between these two VMs.

\subsection{Data exchange cost}

We assume the data exchange cost $c^{exch}_{yy'}$ to be strictly
larger than zero when two connected components are assigned to VMs on different SPs $s_y$ and $s_{y'}$~\cite{19};
otherwise, $c^{exch}_{yy'}=0$. This element captures the cost incurred from traffic
exchange among different SPs in a VC.

\subsection{Problem formulation}

To better analyze the proposed multi-task offloading framework, we mainly
study the problem in one VC because the proposed mechanism is universal across all
VCs. Furthermore, we focus on the snapshot of one VC in the network, that contains $|
\bm{O}|$ TOs, $o_x\in \bm{O}$, and $|\bm{S}|$ SPs, $s_y\in
\bm{S}$, where each $o_x
$ has a graph-based task ${\bm{G}}^{\bm{o_x}}$ waiting
to be offloaded to available VMs in ${\bm{G}}^{\bm{s}}$. Notably,
the time of data transmission and the resulting feedback regarding execution between
TOs and SPs\footnote{Here, it is assumed that the transmission time of
component data and the resulting feedback between TOs to SPs are ignored
for two main reasons~\cite{19}. First, some computation-intensive applications
have small input/output data sizes that can be ignored, such as vehicular
navigation. Second, advanced communication standards and technology can
achieve low end-to-end latency (e.g., 5G)~\cite{17}, which leads to
cases where the data transmission time of each component can be ignored.} are ignored~\cite{17,19}. The notation $s_y\in \mathcal{\bm{R}}_{\bm{o_x}}$, where $o_x
\in \bm{O}$ and $s_y\in \bm{S}$ indicates that $s_y$ is
in $o_x$'s communication range. We only consider one-hop V2V computation offloading
in this paper, meaning that $o_x$ cannot offload any data to $s_y$ when $s_y\notin
\mathcal{\bm{R}}_{\bm{o_x}}$.

Let the binary indicator ${\kappa}^{xy}_{ij}=1$ denote the assignment of component $v^{o_x}_{i}$ of
TO $o_x$ to VM $s^{y}_{j}$ of SP $s_y$; and ${\kappa}^{xy}_{ij}=0$ otherwise.
For the data exchange cost model of task partitioning among different SPs, let the binary
indicator ${\nu}^{yy'}_{jj'}=1$ denote data exchange between $s_y$ and $s_{y'}$ upon allocation of components; otherwise, ${\nu}^{yy'}_{jj'}=0$.

Based on the above notations,  ${\nu}^{yy'}_{jj'}$ is defined as a piecewise
function of ${\kappa}^{xy}_{ij}$ and ${\kappa}^{xy'}_{i'j'}$, given in (1),
describing the case where two connected components of a task assigned to VMs on different
SPs will incur data exchange cost.

\begin{align}
{{\nu}^{yy'}_{jj'}} = \begin{cases}
1, & \forall e_{ii'}^{{o_x}} \in \bm{E^{x_o}},\\
&{\rm if}~ y \ne y'~{\rm and} ~{\kappa}^{xy}_{ij}\times{\kappa}^{xy'}_{i'j'} = 1 \\
0, &\rm otherwise.
\end{cases}
\end{align}

The task completion time is given by the processing time of the slowest processed component
of the task as:
\begin{align}
\bm{U^t_{o_x}}(\bm{\mathcal{K}})=\max {[{\kappa}^{xy}_{ij}\times t^{s_y}_j]}_{1
\le i\le \left|{\bm{n}}_{\bm{x}}\right|,1\le y\le \left|\bm{S}
\right|,1\le j\le \left|{\bm{m}}_{\bm{y}}\right|}
\end{align}
where $\bm{\mathcal{K}}
={\left[{\kappa}^{xy}_{ij}\right]}_{1\le x\le \left|{\bm{O}}\right|,1\le i\le \left|{\bm{n}}_{
\bm{x}}\right|,1\le y\le \left|\bm{S}\right|,1\le j\le \left|{\bm{m}}_{
\bm{y}}\right|}$ defines the matrix of indicator ${\kappa}^{xy}_{ij}$,
for notational simplicity. Then, the total cost of multi-task offloading is formulated
as the following function:
\begin{align}
\bm{U^c}\left(\bm{\mathcal{V}}\left(\bm{\mathcal{K}}\right)\right)={1}/{2}\sum^{\left|\bm{S}
\right|}_{y=1}\sum^{\left|{\bm{m}}_{\bm{y}}\right|}_{j=1}{\sum^{
\left|\bm{S}\right|}_{y'=1}{\sum^{\left|{\bm{m}}_{{\bm{y}}'}\right|}_{j'=1} {{\nu}^{yy'}_{jj'}}}}\times c^{exch}_{yy'}
\end{align}
where $\bm{\mathcal{V}}
\left(\bm{\mathcal{K}}\right)={\left[{\nu}^{yy'}_{jj'}\right]}_{1\le y\le \left|\bm{S}
\right|,1\le j\le \left|{\bm{m}}_{\bm{y}}\right|,1\le y'\le \left|
\bm{S}\right|,1\le j'\le \left|{\bm{m}}_{{\bm{y}}'}
\right|}$ denotes the matrix of indicator ${\nu}^{yy'}_{jj'}$, the elements of which are defined in (1). The normalization factor $1/2$ is considered since data exchange cost will be calculated twice due to the undirected graph-based task model.

Correspondingly, attempting to minimize the completion time of each
task and data exchange costs under opportunistic contact and resource constraints,
we formulate the concurrent multi-task offloading problem under graph-based representation
as shown in (4). For notational simplicity, ${\bm{U}}^{\bm{t}}={[\bm{U^t_{o_x}}
\left(\bm{\mathcal{K}}\right)]}_{1\le x\le |\bm{O}|}$ denotes the vector
of TOs' task completion time.
\begin{align}
\mathop{argmin}_{\bm{\mathcal{K}}} {\xi}_t{\left\|{\bm{U}}^{\bm{t}}
\right\|}_2+{\xi}_c\bm{U^c}(\bm{\mathcal{V}}\left(\bm{\mathcal{K}}\right))
\end{align}
\textrm{s.t.}
\setcounter{equation}{3}
\begin{subequations}
\begin{align}
&\sum^{|\bm{O}|}_{x=1}\sum^{\left|{\bm{n}}_{
\bm{x}}\right|}_{i=1}{{\kappa}^{xy}_{ij}}\le \left|{\bm{m}}_{
\bm{y}}\right|, \forall s_y\in \bm{S}\\
&e^{{-{\lambda}_{yy'}}\times {\omega}^{o_x}_{ii'}}\ge \varepsilon,
\forall e^{o_x}_{ii'}\in {\bm{E}}^{\bm{o_x}},
~y\ne y',{\kappa}^{xy}_{ij}\times {\kappa}^{xy'}_{i'j'}=1\\
&{\kappa}^{xy}_{ij}\triangleq 0, ~{\rm if}~t^{o_x}_i<t^{s_y}_j~{\rm or}~s_y{\notin \mathcal{\bm{R}}_{\bm{o_x}}}
\end{align}
\end{subequations}

\noindent where ${\|\cdot \|}_2$ represents the vector's 2-norm defined as (5), with the significance of minimizing the completion time of each task rather than the overall completion time of all the tasks, in order to better achieve the fairness among TOs:
\begin{align}
{\left\|{\bm{U}}^{\bm{t}}\right\|}_2={\left(\sum^{|\bm{O}|}_{x=1}{{(\bm{{U}^t_{o_x}}
\left(\bm{\mathcal{K}}\right))}^2}\right)}^{\frac{1}{2}}
\end{align}
Here, ${
\xi}_t$ and ${\xi}_c$ are non-negative weight coefficients that indicate the preference
between task completion time and data exchange costs, respectively. The
system prefers to reduce the completion time of tasks with a larger value of ${\xi
}_t$; comparatively, the data exchange cost between vehicles is more likely to be
reduced as the value of ${\xi}_c$ becomes higher. In (4), constraint $(4a
)$ prevents overloading of the available resources, where $\left|{\bm{m}}_{
\bm{y}}\right|$ denotes the number of idle VMs on $s_y$. Constraint $(4b)$ is a probabilistic constraint handling the cases in which components $v^{o_x}_{i}$ and $v^{o_x}_{i'}$ are assigned to VMs on different SPs $s_y$ and $s_{y'}$.
In such cases, the probability of the contact duration between $s_y$ and $s_{y'}$ exceeding ${
\omega} ^{o_x}_{ii'}$ must be greater than the tunable threshold $\varepsilon \in [0,1] $.
Constraint $(4c)$ ensures component $v^{o_x}_{i}$ be
allocated to a VM that meets $v^{o_x}_{i}$'s requirement $ t^{o_x}_i$ to guarantee successful task execution, and one-hop V2V computation offloading from TOs to SPs.

The objective function in (4) represents a nonlinear integer programming problem
that is NP-hard. Moreover, the constraints related to (4) impose solving the sub-graph isomorphism problem, which is NP-complete~\cite{3,18}.
Consequently, the system can rarely identify solutions to reconfigure the IoV extemporaneously,
as the running time required to solve large and real-life network cases increases
sharply with increasing vehicular density (and with the complexity of the VC topology
and task structures). To solve the multi-task offloading problem under graph-based
representation, an optimal algorithm for low-traffic (e.g., fewer than three TOs and
six SPs in a VC) scenarios is first presented. Then, a novel multi-task offloading
algorithm based on connection-restricted random matching is proposed for rush-hour
scenarios, through which near-optimal solutions can be obtained by achieving much
lower computational complexity.

\section{The Optimal multi-task offloading under graph-based representation}

%Algorithm1
\begin{algorithm*}[h!]
\SetKwInOut{Input}{input}\SetKwInOut{Output}{output}
\caption{The Optimal multi-task
offloading algorithm under graph-based representation}
\Input{All graph-based tasks ${\bm{G}}^{\bm{o_x}}$, $x\in  \{1, \dots ,  \left|\bm{O}\right|\}$, VC graph ${\bm{G}}^{\bm{s}}$}
\Output{Optimal solution ${\bm{\mathcal{K}}}^{\bm{*}}$ for distributing all ${\bm{G}}^{\bm{o_x}}$, $x\in  \{1, \dots  ,  \left|\bm{O}\right|\}$ over ${\bm{G}}^{\bm{s}}$}

\textit{// Stage 1 Solution search procedure}  \\
Initialization: $\bm{\mathcal{V}^{*}}\leftarrow \{{\bm{V}}^{\bm{o_1}}\cup {\bm{V}}^{\bm{o_2}}\cup \dots\cup {\bm{V}}^{\bm{o_{|\bm{O}|}}}\}$; $\bm{\mathcal{V}^{**}}\leftarrow \{{\bm{m}}_{\bm{1}}\cup {\bm{m}}_{\bm{2}}\cup \dots  \cup {\bm{m}}_{|\bm{S}|}\}$; $\bm{\mathcal{M}}\leftarrow \emptyset $; ${\bm{\mathcal{K}}}^{\bm{*}}\leftarrow \emptyset $; ~~~~~~~~~~~~~~~~$\%$ Deriving the set of all the permutations of the elements in $\bm{\mathcal{V}^*}$, i.e., $\bm{\mathcal{V}^{comp}}$, and the set of all the permutations of any $|\bm{\mathcal{V}^*}|$ elements in $\bm{\mathcal{V}^{**}}$, i.e., $\bm{\mathcal{V}^{VM}}$\;

${\bm{\mathcal{V}^{comp}}}\bm{\leftarrow} \{{
\bm{V^{c}_i}}|i\bm{\in} \{1,2,\dots, (\sum_{o_x\in\bm O}{|\bm{n_x}|})!\},|\bm{V^{c}_i}|=|\bm{\mathcal{V}^{*}}|$, where each sequence $\bm{V^{c}_i}$ is a permutation of the elements in $\bm{\mathcal{V}^{*}}\}$\;

${\bm{\mathcal{V}^{VM}}}\bm{\leftarrow} \{{
\bm{V^{vm}_j}}|j\bm{\in} \{1,2,\dots, \mathcal{C}(\sum_{s_y\in\bm S}{|\bm{m_y}|}, \sum_{o_x\in\bm O}{|\bm{n_x}|})\},|\bm{V^{vm}_j}|=|\bm{\mathcal{V}^{*}}|$, where each sequence $\bm{V^{vm}_j}$ is a permuatation of any  $|\bm{\mathcal{V}^*}|$ elements in $\bm{\mathcal{V}^{**}}\}$\;

\For{$i=1$ to $(\sum_{o_x\in \bm O}{|\bm{n_x}|})!$}
{\For{$j=1$ to $\mathcal{C}(\sum_{s_y\in\bm S}{|\bm{m_y}|}, \sum_{o_x\in\bm O}{|\bm{n_x}|})$}
{\If{components in $\bm{V^{c}_i}$ can be offloaded to VMs in $\bm{V^{vm}_j}$ sequentially under constraint $(4b)$, $(4c)$ in $(4)$}{$\bm{{\mathcal{K}}_{ij}}\leftarrow {\{\bm{V^{c}_i}, \bm{V^{vm}_j}\}}$; $\%$ match the components in $\bm{V^{c}_i}$ to the VMs in $\bm{V^{vm}_j}$ one by one, as a mapping $\bm{\mathcal{K}_{ij}}$\; $\bm{\mathcal{M}}\leftarrow \bm{\mathcal{M}} \cup \bm{\mathcal{K}_{ij}};$}

{\Else{$\bm{{\mathcal{K}}_{ij}}\leftarrow\emptyset $\;}}}}

\textit{// Stage 2 The optimal solution selection procedure}\\
${\bm{\mathcal{K}}}^{
\bm{*}}\bm{\leftarrow} $ the solution with minimum value of
(4) in $\bm{\mathcal{M}}$\; 
\textbf{end algorithm}
\end{algorithm*}

Our optimal algorithm aims to solve the graph-based multi-task offloading
problem by enumerating all possible solutions. The pseudocode is
given in \textbf{Algorithm I}, where $\bm{V^c_i}$ and $\bm{V^{vm}_j}$ are sequence of permutation of the components and VMs. Notation $\bm{\mathcal{M}}$ and $\bm{\mathcal{K}^*}$ indicate the set of possible solutions and the optimal mapping from components to VMs, respectively.
The algorithm contains two primary stages: in Stage 1, i.e., lines 2--10, we
look for every possible solution $\bm{\mathcal{K}_{ij}}$, which is seen as a mapping from component set to possible VM set, based on merging the adjacency matrices of tasks
into a sparse matrix to handle irrelevant features between these concurrent tasks; in Stage 2, i.e., lines 12--13, the best solution  with the minimum
value of (4) is selected. In this algorithm, going through all possible solutions ensures identification
of the optimal solution for the offloading problem via graph-based representation.

However,
obtaining the optimal solution requires high computational complexity of $\mathcal{O}(
(\sum_{o_x\in{\bm{O}}}{|{\bm{n}}_{\bm{x}}|})!\times \mathcal{C} (\sum_{s_y\in{\bm{S}}}{|{
\bm{m}}_{\bm{y}}|}, $ $\sum_{o_x\in{\bm{O}}}{|{\bm{n}}_{\bm{x}}
|}))$ where $\sum_{o_x\in{\bm{O}}}{|{\bm{n}}_{\bm{x}}|}$ represents
the total number of components of multi-tasks and ``$!$'' is the factorial
notation. $\sum_{s_y\in{\bm{S}}}{|{\bm{m}}_{\bm{y}}|}$ indicates the
number of available VMs in the related VC. Notably, $\mathcal{C} (m, n)$ stands for the $m-choose-n$ operation,
as tasks are modeled as undirected graphs where components do not have a particular
execution sequence. This computational complexity becomes prohibitive as the number
of TOs and SPs and the available VMs grow larger.

%Algorithm2
\begin{algorithm*}[h!]
\SetKwInOut{Input}{input}\SetKwInOut{Output}{output}
\caption{Connection-restricted random matching for
multi-task offloading under graph-based representation}
\Input{Tasks ${\bm{G}}^{\bm{o_x}}$, $x\in \{1, \dots  , \left|\bm{O}\right|\}$, VC graph ${\bm{G}}^{\bm{s}}$, the number of iterations $\gamma$}
\Output{Near-optimal
solution ${\bm{\mathcal{K}}}^{\bm{*}}$ for distributing all ${\bm{G}}^{\bm{o_x}}$, $x\in \{1,\dots,  \left|\bm{O}\right|\}$ over ${\bm{G}}^{\bm{s}}$}

\textit{// Stage 1 Initialization}

$\bm{\mathcal{V}^{comp}}\leftarrow \{{\bm{V}}^{\bm{o_1}}\cup {\bm{V}}^{\bm{o_2}}\cup \dots\cup {\bm{V}}^{\bm{o_{|\bm{O}|}}}\}$; $\bm{\mathcal{V}^{VM}}\leftarrow \{{\bm{m}}_{\bm{1}}\cup {\bm{m}}_{\bm{2}}\cup \dots  \cup {\bm{m}}_{|\bm{S}|}\}$;

${\bm{\mathcal{K}_0}}\leftarrow \emptyset $; ${\bm{\mathcal{K}}}^{\bm{*}}
\leftarrow \emptyset $; $\vartheta_0=+\infty $;

\textit{// Stage 2 Connection-restricted random matching}

\If{$\bm{\mathcal{V}^{VM}}=\emptyset$}{
go to Step 28;}

\For{$r=1$ to $\gamma$}{
	 $k\leftarrow 1$;\\
	\If{$k\le \left|\bm{\mathcal{V}^{comp}}\right|$}{
		randomly select a component $v_k\in \bm{\mathcal{V}^{comp}}, $ and
		assign $v_k$ uniformly at random to an available VM denoted as $s'_k
		\in \bm{\mathcal{V}^{VM}}$;

		$k\leftarrow k+1$; $\bm{\mathcal{V}^{comp}}\leftarrow \bm{\mathcal{V}^{comp}} \backslash {\{v}_k\}$; $\bm{\mathcal{V}^{VM}}
		\leftarrow \bm{\mathcal{V}^{VM}}\backslash {\{s}'_k\}$; ${\bm{\mathcal{K}_{r}}}\leftarrow \{(v_k, s'_k) \}$;

		\For{$k\le \left|\bm{\mathcal{V}^{comp}} \right|$}{
			randomly select a component denoted as $v_k\in \bm{\mathcal{V}^{comp}}$\;
			\If{there exist edges between $v_k$ and components in ${\bm{\mathcal{K}_r}}$, and available VMs,}{
			assign $v_k$ uniformly at random to an available VM denoted
			as $s'_k\in \bm{\mathcal{V}^{VM}}$ while meeting constraints $(4b) $ and $(4c)$ in (4)\;
			}
			\ElseIf{there exist edges between $v_k$ and components in ${\bm{\mathcal{K}_r}}$, and no available VMs,}{
			${\bm{\mathcal{K}_r}}\leftarrow {\bm{\mathcal{K}_{r-1}}}$; go to Step 26\;
			}
			\ElseIf{there are no edges between $v_k$ and components in ${\bm{\mathcal{K}_r}}$, and there exist available VMs,}{
			assign $v_k$ uniformly at random to an available VM denoted
			as $s'_k\in \bm{\mathcal{V}^{VM}}$ while meeting constraints $(4c)$ in (4)\;
			}
			\Else{${\bm{\mathcal{K}_r}}
			\leftarrow {\bm{\mathcal{K}_{r-1}}}$; go to Step 26;}

			$\bm{\mathcal{V}^{comp}}\leftarrow \bm{\mathcal{V}^{comp}}
			\backslash {\{v}_k\}$; $\bm{\mathcal{V}^{VM}}{\mathrm{\leftarrow
			} }\bm{\mathcal{V}^{VM}}\backslash {\{s'_k}\}$; ${\bm{\mathcal{K}_r}}
			\leftarrow {\bm{\mathcal{K}_r}}\bm{
			\cup} \{(v_k, s'_k)\}$; $k
			\leftarrow k+1$;
		}
	}
	$\vartheta_{r}\leftarrow $ value of
	(4) on solution ${\bm{\mathcal{K}_r}}$;\\
	\If{$\vartheta_{r}\ge \vartheta_{r-1}$}{
	${\bm{\mathcal{K}_r}}\leftarrow {\bm{\mathcal{K}_{r-1}}}$;
	}
	$r\leftarrow r+1$;
}

${\bm{\mathcal{K}}}^{\bm{*}}\leftarrow {\bm{\mathcal{K}_r}}$;

\textbf{end algorithm}

\end{algorithm*}

\section{Multi-task offloading based on connection-restricted random matching}

To better handle the multi-task offloading problem under graph-based representation
in real-life IoV scenarios with large numbers of SPs and TOs as well as complicated
structures, we develop an efficient low-complexity multi-task offloading algorithm
called connection-restricted random matching (CRRM). Through the proposed CRRM algorithm,
we have only one visit to each component of tasks, for which an available VM can
be randomly chosen for execution while satisfying the inner structures of tasks and
the VC, modeled as graphs. Consequently, the computational complexity of $\mathcal{O}(\sum_{o_x\in{\bm{O}}}{
\left|{\bm{n}}_{\bm{x}}\right|})$ in each iteration can be achieved
to substantially improve the running time performance. Details appear in \textbf{Algorithm II}, where $\bm{\mathcal{K}_r}$ and $\bm{\mathcal{K}^*}$ indicates the mapping between the components and VMs in iteration $r$ and the near-optimal solution, respectively.

In
CRRM, Steps 2--3 describe the initialization procedure, and Steps 5--6 handle the unsuccessful offloading cases
due to insufficient available VMs. Steps 7--22 describe how each component is matched to an available VM. Specifically,
Steps 14--15 and 18--19 indicate that a component is randomly mapped to an available
VM while satisfying V2V communication coverage and required execution time limitations
as well as opportunistic connection restrictions between SPs. Notably, potential
competition on VMs between components exists during matching; thus, Steps 16--17
and 20--21 deal with unsuccessful offloading cases in each iteration, where no available VMs can be used to handle the randomly selected component. A better solution is considered after every iteration by comparing the
value of the objective function given in (4), among which the best solution will
be reserved as the final solution $\bm{\mathcal{K}^*}$  through Steps 23--27. This process ensures convergence
of the proposed algorithm.

\section{Numerical results and performance evaluation}

This section presents numerical results illustrating the performance of
the proposed algorithms. Moreover, two greedy-based algorithms serve as baseline
methods are shown below to better evaluate the advantages of the optimal and proposed
CRRM algorithms.

\noindent \textbf{Degree preferred mechanism (DPM):} Components are sorted by degree\footnote{The degree of a vertex in a graph is defined as the number of edges
associated with this vertex.} from largest to smallest as a list. Each component
in the list is matched one by one to the available VM with the largest degree at
present, while satisfying all the constraints in (4), until all components have been
allocated successfully. Otherwise, tasks must be executed locally.

\noindent \textbf{Execution time preferred mechanism (ETPM):} Randomly select one
task component at a time and match it to the available VM that can provide the
lowest execution time at present, while satisfying all the constraints in (4), until
all components have been allocated successfully. Otherwise, tasks must be executed
locally.

The completion time of a locally executed task depends on the number and
execution time of available VMs the TO can provide, which is considered a serial
mode rather than parallel, and thus excluded from the solution space of the proposed
optimal and CRRM algorithms. Thus, local execution cases are not considered in our
performance comparisons and evaluations. In this paper, task types are randomly chosen from those depicted in Fig. 1(b) and the related parameters are randomly set in the following intervals: $\varepsilon \in [0.9,1)$, ${\omega}^{o_x}_{ii'}\in [0.1,0.3]$, $t^{o_x}_i\in [0.1,0.2]$, $t^{s_y}_j\in [0.05,0.25
]$, $c^{exch}_{yy'}\in [0.05,0.15]$, ${\xi}_t={\xi}_c=0.5$, ${\lambda}_{yy'}\in [0.04,0.05]$ for
low-traffic cases and ${\lambda}_{yy'}\in [0.01,0.02]$ for rush-hour cases.

%f2
\begin{figure}[h!t]
\centerline{\includegraphics[width=3.0in,height=4.0cm]{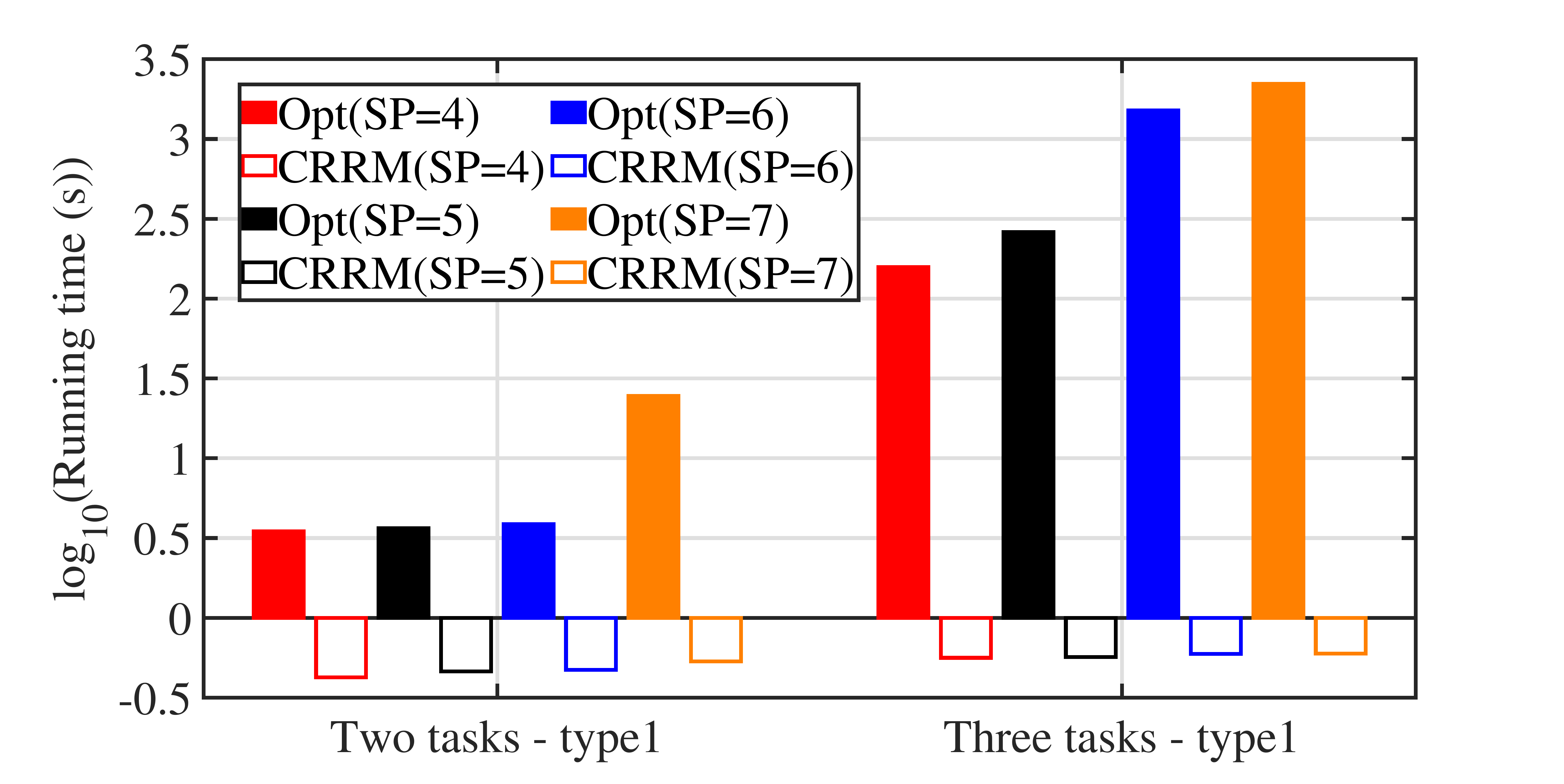}}
\caption{Running time comparison of Task type 1 between the optimal and CRRM
algorithms.}
\label{fig2}
\end{figure}

%t1
\begin{table}
\centering
\begin{center}\renewcommand\arraystretch{1.2} \caption{Comparison on Running time performance  between the optimal algorithm and CRRM under different number of iterations}
\setlength{\tabcolsep}{0.5mm}{
\begin{tabular}{|c|c|c|c|c|}
\hline
\multirow{2}*{Tasks, SPs, and VMs in a VC} & \multirow{2}*{Optimal} & \multicolumn{3}{c|}{Iterations (CRRM)}\\
\cline{3-5}
&   & 1,000& 2,000& 3,000\\
\hline
2 $\times$ type 2,~~4 $\times$ SPs,~~16 $\times$ VMs & $\geq5\times10^3$s  &0.1413s & 0.2816s & 0.4194s \\
5 $\times$ type 2,~~8 $\times$ SPs,~~40 $\times$ VMs &  $\geq1\times10^4$s   &0.1439s & 0.2745s & 0.4275s \\
2 $\times$ type 3,~~4 $\times$ SPs,~~25 $\times$ VMs  & $\geq1\times10^4$s  &0.1613s & 0.3120s & 0.4806s \\
5 $\times$ type 3,~~8 $\times$ SPs,~~48 $\times$ VMs & $\geq2\times10^4$s   &0.1702s & 0.3442s & 0.5103s \\
2 $\times$ type 4,~~5 $\times$ SPs,~~30 $\times$ VMs & $\geq3\times10^4$s &0.1688s & 0.3234s & 0.4956s \\
5 $\times$ type 4,~~10 $\times$ SPs,~~50 $\times$ VMs &$\geq3\times10^4$s  &0.1801s & 0.3628s & 0.5279s \\
\hline
\end{tabular}}
\end{center}
\end{table}

%f3
\begin{figure*}[t]\centering	
    \subfigure[]{{\includegraphics[width=1.74in,height=2.6cm]{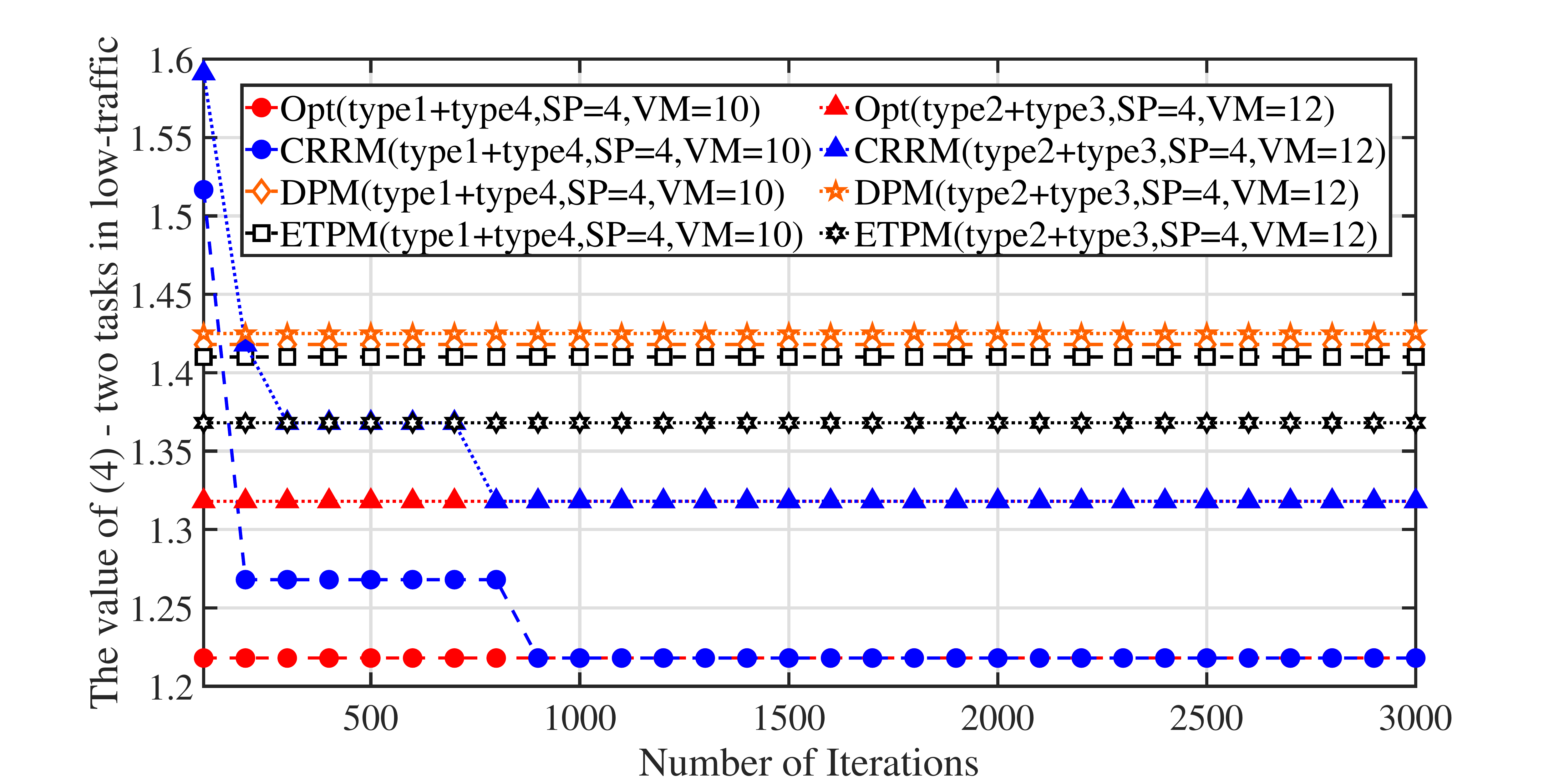}}}
    \subfigure[]{{\includegraphics[width=1.74in,height=2.6cm]{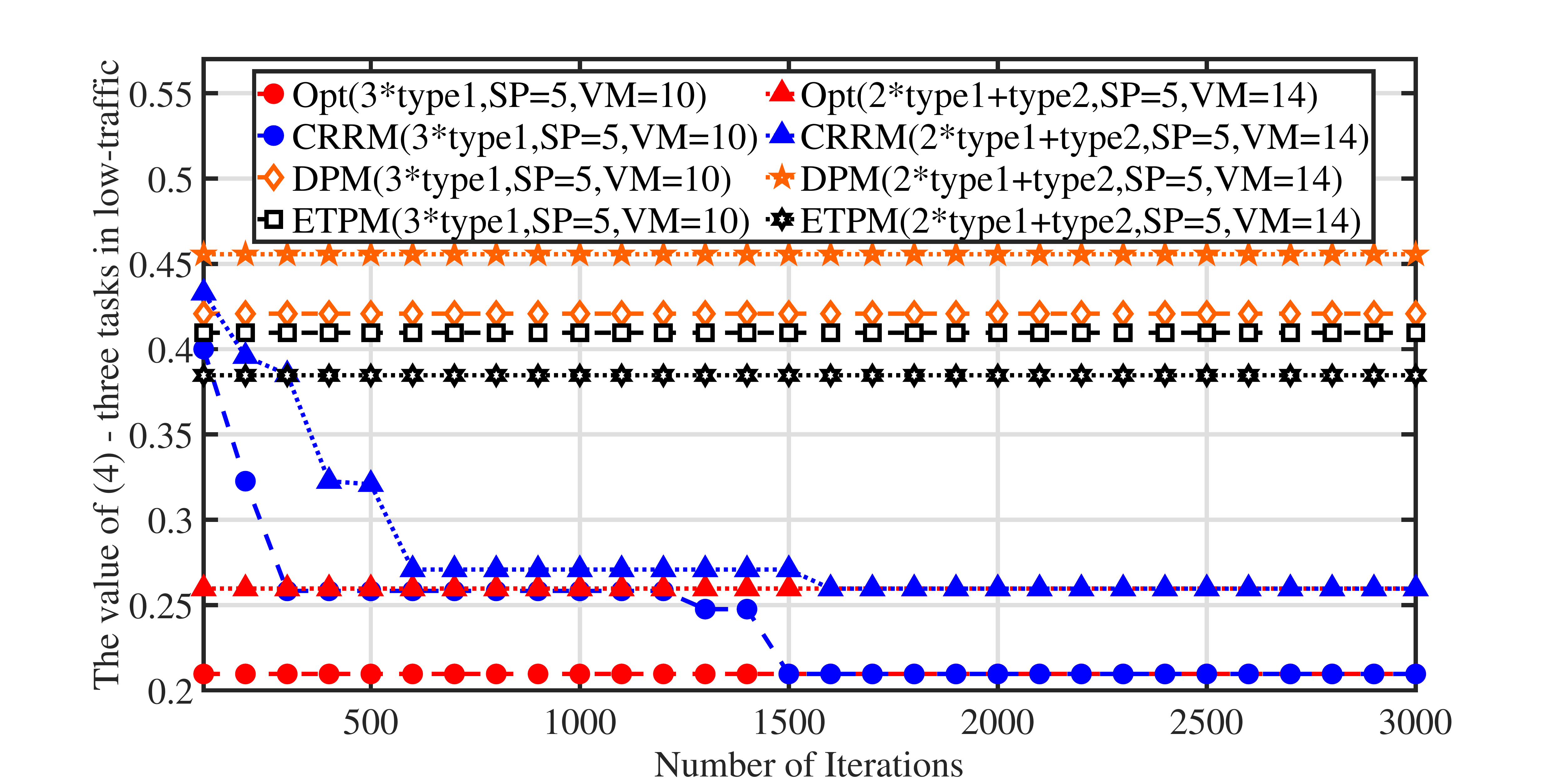}}}
    \subfigure[]{{\includegraphics[width=1.74in,height=2.6cm]{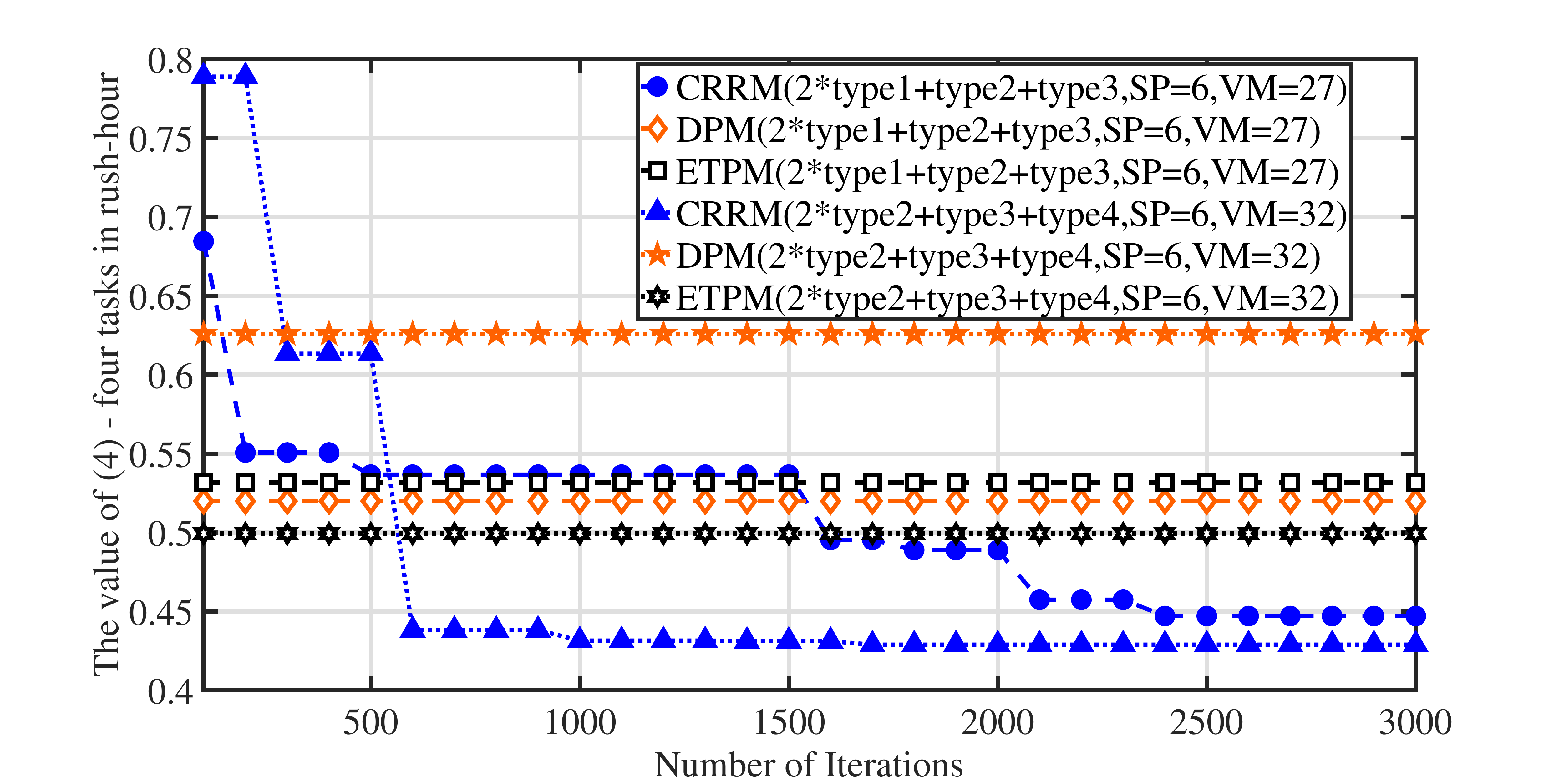}}}
    \subfigure[]{{\includegraphics[width=1.74in,height=2.6cm]{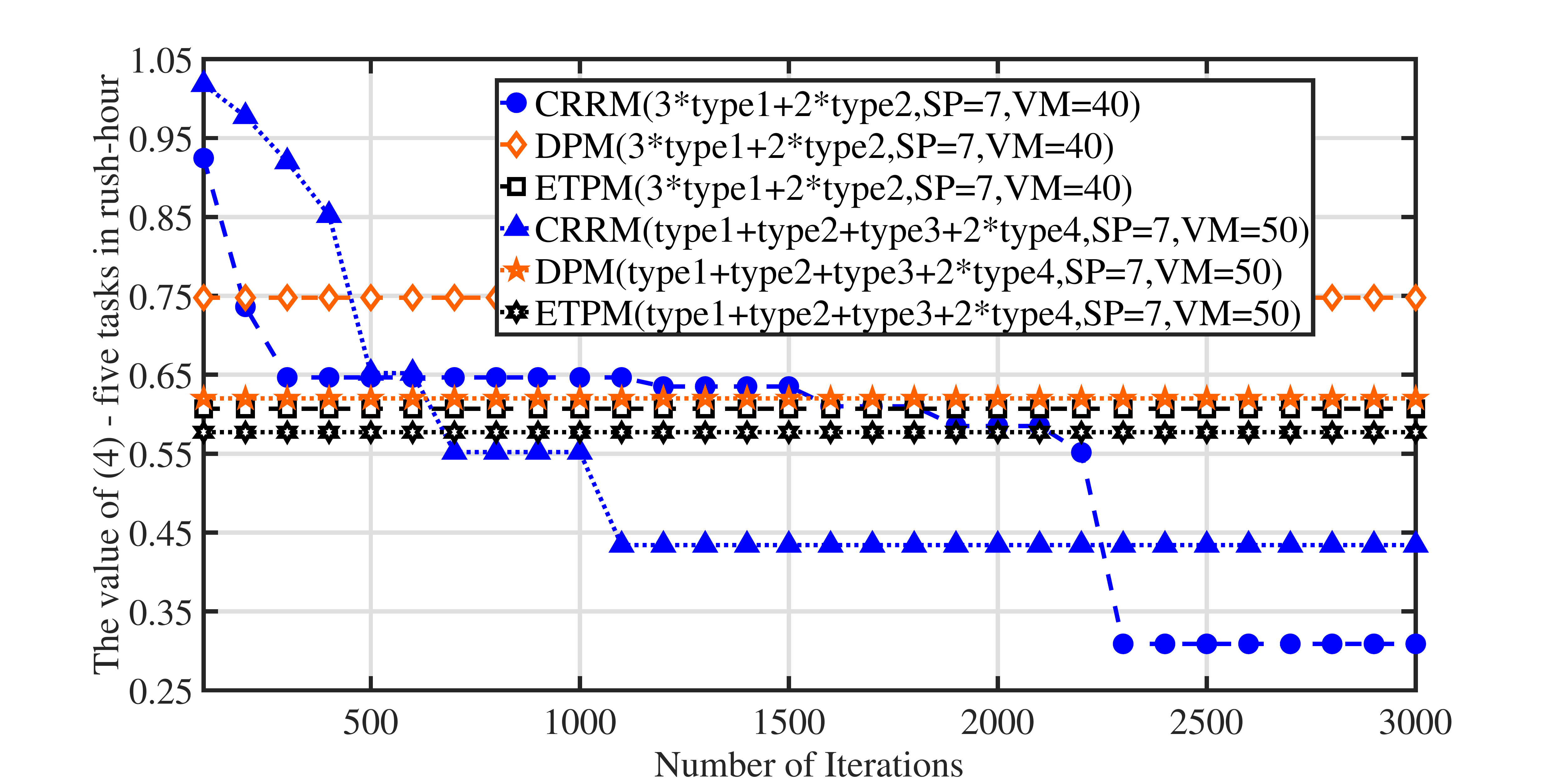}}}
\caption{Performance comparison of the value of (4) in different scenarios: a) VCs containing two concurrent tasks and four SPs in low-traffic; b) VCs containing
three concurrent tasks and five SPs in low-traffic; c) VCs containing four concurrent tasks and six SPs in rush-hour; d)  VCs containing
five concurrent tasks and seven SPs in rush hour.}
\label{fig3}
\end{figure*}
Due to that the running time can be too large as the task structure becomes
more complex, we mainly consider different numbers of Task type 1 in various scenarios,
for which a running time comparison between the optimal and CRRM algorithms appears
in Fig. 2 under 10-based logarithm presentation. Compared with the CRRM algorithm
under 3,000 iterations, Fig. 2 indicates that the running time of the optimal algorithm
may rise sharply as the vehicular density grows (e.g., approximately 2,261 seconds (37.68 minutes) are needed to obtain the optimal offloading solution in a VC containing seven SPs
and three tasks), which makes it unsuitable for fast-changing and large IoV networks.
Notably, VC configurations with different topological complexity (e.g., the presence
of more connections between vehicles) will also lead to dramatic changes in running
time performance. However, Fig. 2 and \textbf{Table I} reveal that the proposed CRRM
algorithm can always maintain a low running time under different numbers of iterations,
enabling which to be implemented efficiently in rapidly changing and large-scale
IoV networks particularly during rush hour, rather than the lowerbound on running time of the optimal algorithm. Considering in Fig. 2, comparing with the optimal algorithm, utilizing the proposed CRRM algorithm leads to 90.19$\%$ and 99.84$\%$ performance improvement on the average running time in scenarios with two tasks and three tasks, respectively.

Performance comparisons on the value of (4) between baseline methods and
the proposed algorithms in low-traffic and rush-hour scenarios with different numbers
of tasks, SPs, and available VMs are presented in Fig. 3.
As the number of iterations increases, the proposed CRRM outperforms the DPM and
ETPM methods. Although these methods may perform similarly initially, when the number
of iterations is small (e.g., Fig. 3(c)), the gap becomes larger as iterations increase.
Due to that a very long time is needed to obtain optimal solutions in rush-hour scenarios,
the performance of the optimal algorithm is ignored in Fig. 3(c) and  Fig. 3(d). Solutions of the proposed CRRM approach those of the optimal algorithm shown in Fig. 3 while enjoying much
lower computational complexity for different task types and vehicular densities.

%%f4
%\begin{figure}[htb!]
%\centering
%\subfigure[]{\includegraphics[width=.95\linewidth]{fig4a}}
%\subfigure[]{\includegraphics[width=.95\linewidth]{fig4b}}
%\caption{Performance comparison of the value of (4) in low-traffic
%scenarios: (a) VCs containing two concurrent tasks and four SPs; (b) VCs containing
%three concurrent tasks and five SPs.}
%\label{fig4}
%\end{figure}
%
%
%
%%f5
%\begin{figure}[h!t]
%\centering
%\subfigure[]{\includegraphics[width=.95\linewidth]{fig5a}}
%\subfigure[]{\includegraphics[width=.95\linewidth]{fig5b}}
%\caption{Performance comparison of the value of (4) in rush-hour
%scenarios: (a) VCs containing four concurrent tasks and six SPs; (b) VCs containing
%five concurrent tasks and seven SPs.}
%\label{fig5}
%\end{figure}

\section{Conclusion}

This paper studies a novel multi-task offloading mechanism over VCs using graph-based
representation, which is modeled as a nonlinear integer programming problem under constraints. For low-traffic scenarios, an optimal algorithm is introduced; for rush-hour
scenarios, a CRRM algorithm is proposed with low computational complexity. The effectiveness
of the proposed algorithms is revealed through comprehensive simulations. One
potential future direction for research is considering tasks modeled by directed graphs.


\begin{thebibliography}{00}
\bibitem{1} G. Qiao, S. Leng, K. Zhang, and Y. He, ``Collaborative Task Offloading
in Vehicular Edge Multi-Access Networks," \textit{IEEE Commun. Mag.}, vol. 56, no. 8,
pp. 48--54, 2018.

\bibitem{2} M. Jia, J. Cao, and L. Yang, ``Heuristic Offloading of Concurrent Tasks
for Computation Intensive Applications in Mobile Cloud Computing,'' \textit{IEEE
Int. Conf Comp. Commun. Workshops (INFOCOM WKSHPS)}, Toronto, CA, Apr. 2014, pp.
352--357.

\bibitem{3} S. Hosseinalipour, A. Nayak, and H. Dai, ``Power-Aware Allocation of Graph Jobs in Geo-Distributed Cloud Networks,'' \textit{IEEE Trans. Parallel Distrib. Syst.}, DOI: 10.1109/TPDS.2019.2943457, pp. 1--1, 2019.

\bibitem{4} J. Ghaderi, S. Shakkottai, and R. Srikant, ``Scheduling Storms and Streams
in The Cloud,'' \textit{ACM Trans. Modeling and Performance Eval. of Comput. Syst.},
vol. 1, no. 4, pp. 1--14, 2016.

\bibitem{5} T. Mekki, I. Jabri, A. Rachedi, and M. B. Jemaa, ``Vehicular Cloud Networks:
Challenges, Architectures, and Future Directions,'' \textit{Veh. Commun.}, vol. 9,
pp. 268--280, 2017.

\bibitem{6} L. Li, Z. Kuang, and A. Liu, ``Energy Efficient and Low Delay Partial
Offloading Scheduling and Power Allocation for MEC,'' \textit{IEEE Int. Conf Commun.
(ICC)}, Shanghai, China, May. 2019, pp. 1--6.

\bibitem{7} M. Chen, B. Liang, and M. Dong, ``Multi-User Multi-Task Offloading and
Resource Allocation in Mobile Cloud Systems,'' \textit{IEEE Trans. Wireless Commun.},
vol. 17, no. 10, pp. 6790--6805, 2018.

\bibitem{8} Y. Liu, S. Wang, J. Huang, and F. Yang, ``A Computation Offloading Algorithm
Based on Game Theory for Vehicular Edge Networks,'' \textit{IEEE Int. Conf. Commun.
(ICC)}, Kansas City, USA, May. 2018, pp. 1--6.

\bibitem{9} Y. Dai, D. Xu, S. Maharjan, and Y. Zhang, ``Joint Computation Offloading
and User Association in Multi-Task Mobile Edge Computing,'' \textit{IEEE Trans. Veh.
Technol.}, vol. 67, no. 12, pp: 12313--12325, 2018.

\bibitem{10} D. Huang, P. Wang, and D. Niyato, ``A Dynamic Offloading Algorithm
for Mobile Computing,'' \textit{IEEE Trans. Wireless Commun.}, vol.11, no. 6, pp.
1991--1995, 2012.

\bibitem{11} L. Shi, Z. Zhang, and T. Robertazzi, ``Energy-Aware Scheduling of Embarrassingly
Parallel Jobs and Resource Allocation in Cloud,'' \textit{IEEE Trans. Parallel Distrib.
Syst.}, vol. 28, no. 6, pp. 1607--1620, 2017.

\bibitem{12} M. Goudarzi, M. Zamani, and A. T. Haghighat ``A Fast Hybrid Multi-Site
Computation Offloading for Mobile Cloud Computing,'' \textit{Journal of Netw. Comput.
Appl.}, vol. 66, pp. 219--231, 2017.

\bibitem{13} F. Sun, F. Hou, N. Cheng, M. Wang, H. Zhou, L. Gui, and X. Shen, ``Cooperative
Task Scheduling for Computation Offloading in Vehicular Cloud,'' \textit{IEEE Trans.
Veh. Technol.}, vol. 67, no. 11, pp: 11049--11061, 2018.

\bibitem{14} Z. Tao, Q. Xia, Z. Hao, C. Li, L. Ma, S. Yi, and Q. Li, ``A Survey of
Virtual Machine Management in Edge Computing,'' \textit{Proc. IEEE}, vol. 107, no.
8, pp: 1482--1499, 2019.

\bibitem{15} X. Zhu, Y. Li, D. Jin, and J. Lu, ``Contact-Aware Optimal Resource
Allocation for Mobile Data Offloading in Opportunistic Vehicular Networks,'' \textit{IEEE
Trans. Veh. Technol.}, vol. 66, no. 8, pp. 7384--7399, 2017.

\bibitem{16} H. Zhu, L. Fu, G. Xue, Y. Zhu, M. Li, and L. M. Ni, ``Recognizing Exponential
Inter-Contact Time in VANETs,'' \textit{IEEE Int. Conf. Comp. Commun. (INFOCOM)},
San Diego, CA, Mar. 2010, pp. 1--5.

\bibitem{17} N. Alliance, ``NGMN 5G White Paper v1. 0'' \textit{approved and delivered
by the NGMN board}, Feb. 2015.

\bibitem{18} V. Carletti, P. Foggia, A. Saggese, and M. Vento, ``Challenging the
Time Complexity of Exact Subgraph Isomorphism for Huge and Dense Graphs with VF3,'' \textit{IEEE
Trans. Pattern Anal. Mach. Intell.}, vol. 40, no. 4, pp. 804--818.

\bibitem{19} M. LiWang, S. Hosseinalipour, Z. Gao, Y. Tang, L. Huang, and H. Dai,
``Allocation of Computation-Intensive Graph Jobs over Vehicular Clouds in IoV,'' \textit{IEEE Internet Things J.}, DOI: 10.1109/JIOT.2019.2949602, pp. 1--1, 2019.
%
%\bibitem{22} Z . Ning, P. Dong, X. Kong and F. Xia, ``A cooperative partial computation offloading
%scheme for mobile edge computing enabled Internet of Thing,'' \textit{IEEE Internet Things
%J.}, vol. 6, no. 3, 2019, pp. 4804--4814.

\end{thebibliography}
\end{document}